\documentclass[aps,prl,article,twocolumn,reprint,superscriptaddress,floatfix,nobibnotes]{revtex4-2}
\usepackage{graphicx,amsfonts,color,comment,amsmath,amssymb,float,hyperref,appendix,dcolumn}
\usepackage{bm}
\usepackage{url}
\usepackage{multirow}
\usepackage{soul}
\preprint{APS/123-QED}
\begin{document}

\title{An Absence of TeV halos around Millisecond Pulsars}

\author{A.U.~Abeysekara}
\affiliation{Department of Physics and Astronomy, University of Utah, Salt Lake City, UT, USA}

\author{R.~Alfaro}
\affiliation{Instituto de F\'{i}sica, Universidad Nacional Autónoma de México, Ciudad de Mexico, Mexico}

\author{C.~Alvarez}
\affiliation{Universidad Autónoma de Chiapas, Tuxtla Gutiérrez, Chiapas, México}

\author{J.C.~Arteaga-Velázquez}
\affiliation{Universidad Michoacana de San Nicolás de Hidalgo, Morelia, Mexico}

\author{D.~Avila Rojas}
\affiliation{Instituto de F\'{i}sica, Universidad Nacional Autónoma de México, Ciudad de Mexico, Mexicosica, Universidad Nacional Autónoma de México, Ciudad de Mexico, Mexico}

\author{H.A.~Ayala Solares}
\affiliation{Department of Physics, Pennsylvania State University, University Park, PA, USA}

\author{R.~Babu}
\affiliation{Department of Physics and Astronomy, Michigan State University, East Lansing, MI, USA}

\author{E.~Belmont-Moreno}
\affiliation{Instituto de F\'{i}sica, Universidad Nacional Autónoma de México, Ciudad de Mexico, Mexico}

\author{A.~Bernal}
\affiliation{Instituto de Astronom\'{i}a, Universidad Nacional Autónoma de México, Ciudad de Mexico, Mexico}

\author{K.S.~Caballero-Mora}
\affiliation{Universidad Autónoma de Chiapas, Tuxtla Gutiérrez, Chiapas, México}

\author{T.~Capistrán}
\affiliation{Instituto de Astronom\'{i}a, Universidad Nacional Autónoma de México, Ciudad de Mexico, Mexico}

\author{A.~Carramiñana}
\affiliation{Instituto Nacional de Astrof\'{i}sica, Óptica y Electrónica, Puebla, Mexico}

\author{S.~Casanova}
\affiliation{Institute of Nuclear Physics Polish Academy of Sciences, PL-31342 IFJ-PAN, Krakow, Poland}

\author{U.~Cotti}
\affiliation{Universidad Michoacana de San Nicolás de Hidalgo, Morelia, Mexico}

\author{J.~Cotzomi}
\affiliation{Facultad de Ciencias F\'{i}sico Matemáticas, Benemérita Universidad Autónoma de Puebla, Puebla, Mexico}

\author{S.~Coutiño de León}
\affiliation{Department of Physics, Wisconsin IceCube Particle Astrophysics Center, University of Wisconsin-Madison, Madison, WI, USA}
\affiliation{\href{mailto:scoutino@ific.uv.es}{scoutino@ific.uv.es}}

\author{E.~De la Fuente}
\affiliation{Departamento de F\'{i}sica, Centro Universitario de Ciencias Exactase Ingenierias, Universidad de Guadalajara, Guadalajara, Mexico}

\author{D.~Depaoli}
\affiliation{Max-Planck Institute for Nuclear Physics, 69117 Heidelberg, Germany}

\author{P.~Desiati}
\affiliation{Department of Physics, University of Wisconsin-Madison, Madison, WI, USA}

\author{N.~Di Lalla}
\affiliation{Department of Physics, Stanford University: Stanford, CA 94305–4060, USA}

\author{R.~Diaz Hernandez}
\affiliation{Instituto Nacional de Astrof\'{i}sica, Óptica y Electrónica, Puebla, Mexico}

\author{M.A.~DuVernois}
\affiliation{Department of Physics, University of Wisconsin-Madison, Madison, WI, USA}

\author{J.C.~Díaz-Vélez}
\affiliation{Department of Physics, University of Wisconsin-Madison, Madison, WI, USA}

\author{K.~Engel}
\affiliation{Department of Physics, University of Maryland, College Park, MD, USA}

\author{T.~Ergin}
\affiliation{Department of Physics and Astronomy, Michigan State University, East Lansing, MI, USA}

\author{K.L.~Fan}
\affiliation{Department of Physics, University of Maryland, College Park, MD, USA}

\author{K.~Fang}
\affiliation{Department of Physics, Wisconsin IceCube Particle Astrophysics Center, University of Wisconsin-Madison, Madison, WI, USA}
\affiliation{\href{mailto:kefang@physics.wisc.edu}{kefang@physics.wisc.edu}}

\author{N.~Fraija}
\affiliation{Instituto de Astronom\'{i}a, Universidad Nacional Autónoma de México, Ciudad de Mexico, Mexico}

\author{S.~Fraija}
\affiliation{Instituto de Astronom\'{i}a, Universidad Nacional Autónoma de México, Ciudad de Mexico, Mexico}

\author{J.A.~García-González}
\affiliation{Tecnologico de Monterrey, Escuela de Ingenier\'{i}a y Ciencias, Ave. Eugenio Garza Sada 2501, Monterrey, N.L., Mexico}

\author{F.~Garfias}
\affiliation{Instituto de Astronom\'{i}a, Universidad Nacional Autónoma de México, Ciudad de Mexico, Mexico}

\author{M.M.~González}
\affiliation{Instituto de Astronom\'{i}a, Universidad Nacional Autónoma de México, Ciudad de Mexico, Mexico}

\author{J.A.~Goodman}
\affiliation{Department of Physics, University of Maryland, College Park, MD, USA}

\author{J.P.~Harding}
\affiliation{Los Alamos National Laboratory, Los Alamos, NM, USA}

\author{S.~Hernández-Cadena}

\author{D.~Huang}
\affiliation{Department of Physics, University of Maryland, College Park, MD, USA}

\author{F.~Hueyotl-Zahuantitla}
\affiliation{Universidad Autónoma de Chiapas, Tuxtla Gutiérrez, Chiapas, México}

\author{A.~Iriarte}
\affiliation{Instituto de Astronom\'{i}a, Universidad Nacional Autónoma de México, Ciudad de Mexico, Mexico}

\author{S.~Kaufmann}
\affiliation{Universidad Politecnica de Pachuca, Pachuca, Hgo, Mexico}

\author{D.~Kieda}
\affiliation{Department of Physics and Astronomy, University of Utah, Salt Lake City, UT, USA}

\author{A.~Lara}
\affiliation{Instituto de Geof\'{i}sica, Universidad Nacional Autónoma de México, Ciudad de Mexico, Mexico}

\author{J.~Lee}
\affiliation{University of Seoul, Seoul, Rep. of Korea}

\author{H.~León Vargas}
\affiliation{Instituto de F\'{i}sica, Universidad Nacional Autónoma de México, Ciudad de Mexico, Mexico}

\author{J.T.~Linnemann}
\affiliation{Department of Physics and Astronomy, Michigan State University, East Lansing, MI, USA}

\author{A.L.~Longinotti}
\affiliation{Instituto de Astronom\'{i}a, Universidad Nacional Autónoma de México, Ciudad de Mexico, Mexico}

\author{G.~Luis-Raya}
\affiliation{Universidad Politecnica de Pachuca, Pachuca, Hgo, Mexico}

\author{K.~Malone}
\affiliation{Los Alamos National Laboratory, Los Alamos, NM, USA}

\author{O.~Martinez}
\affiliation{Facultad de Ciencias F\'{i}sico Matemáticas, Benemérita Universidad Autónoma de Puebla, Puebla, Mexico}

\author{J.~Martínez-Castro}
\affiliation{Centro de Investigaci\'on en Computaci\'on, Instituto Polit\'ecnico Nacional, M\'exico City, M\'exico.}

\author{J.A.~Matthews}
\affiliation{Dept of Physics and Astronomy, University of New Mexico, Albuquerque, NM, USA}

\author{P.~Miranda-Romagnoli}
\affiliation{Universidad Autónoma del Estado de Hidalgo, Pachuca, Mexico}

\author{J.A.~Morales-Soto}
\affiliation{Universidad Michoacana de San Nicolás de Hidalgo, Morelia, Mexico}

\author{E.~Moreno}
\affiliation{Facultad de Ciencias F\'{i}sico Matemáticas, Benemérita Universidad Autónoma de Puebla, Puebla, Mexico}

\author{M.~Mostafá}
\affiliation{Department of Physics, Temple University, Philadelphia, Pennsylvania, USA}

\author{M.~Najafi}
\affiliation{Department of Physics, Michigan Technological University, Houghton, MI, USA}

\author{L.~Nellen}
\affiliation{Instituto de Ciencias Nucleares, Universidad Nacional Autónoma de Mexico, Ciudad de Mexico, Mexico}

\author{M.~Newbold}
\affiliation{Department of Physics and Astronomy, University of Utah, Salt Lake City, UT, USA}

\author{M.U.~Nisa}
\affiliation{Department of Physics and Astronomy, Michigan State University, East Lansing, MI, USA}

\author{R.~Noriega-Papaqui}
\affiliation{Universidad Autónoma del Estado de Hidalgo, Pachuca, Mexico}

\author{N.~Omodei}
\affiliation{Department of Physics, Stanford University: Stanford, CA 94305–4060, USA}

\author{Y.~Pérez Araujo}
\affiliation{Instituto de F\'{i}sica, Universidad Nacional Autónoma de México, Ciudad de Mexico, Mexico}

\author{E.G.~Pérez-Pérez}
\affiliation{Universidad Politecnica de Pachuca, Pachuca, Hgo, Mexico}

\author{C.D.~Rho}
\affiliation{Department of Physics, Sungkyunkwan University, Suwon 16419, South Korea}

\author{D.~Rosa-González}
\affiliation{Instituto Nacional de Astrof\'{i}sica, Óptica y Electrónica, Puebla, Mexico}

\author{E.~Ruiz-Velasco}
\affiliation{Max-Planck Institute for Nuclear Physics, 69117 Heidelberg, Germany}

\author{M.A.~Roth}
\affiliation{Los Alamos National Laboratory, Los Alamos, NM, USA}

\author{H.~Salazar}
\affiliation{Facultad de Ciencias F\'{i}sico Matemáticas, Benemérita Universidad Autónoma de Puebla, Puebla, Mexico}

\author{A.~Sandoval}
\affiliation{Instituto de F\'{i}sica, Universidad Nacional Autónoma de México, Ciudad de Mexico, Mexico}

\author{J.~Serna-Franco}
\affiliation{Instituto de F\'{i}sica, Universidad Nacional Autónoma de México, Ciudad de Mexico, Mexico}

\author{Y.~Son}
\affiliation{University of Seoul, Seoul, Rep. of Korea}

\author{R.W.~Springer}
\affiliation{Department of Physics and Astronomy, University of Utah, Salt Lake City, UT, USA}

\author{O.~Tibolla}
\affiliation{Universidad Politecnica de Pachuca, Pachuca, Hgo, Mexico}

\author{K.~Tollefson}
\affiliation{Department of Physics and Astronomy, Michigan State University, East Lansing, MI, USA}

\author{I.~Torres}
\affiliation{Instituto Nacional de Astrof\'{i}sica, Óptica y Electrónica, Puebla, Mexico}

\author{R.~Torres-Escobedo}

\author{R.~Turner}
\affiliation{Department of Physics, Michigan Technological University, Houghton, MI, USA}

\author{F.~Ureña-Mena}
\affiliation{Instituto Nacional de Astrof\'{i}sica, Óptica y Electrónica, Puebla, Mexico}

\author{E.~Varela}
\affiliation{Facultad de Ciencias F\'{i}sico Matemáticas, Benemérita Universidad Autónoma de Puebla, Puebla, Mexico}

\author{L.~Villaseñor}
\affiliation{Facultad de Ciencias F\'{i}sico Matemáticas, Benemérita Universidad Autónoma de Puebla, Puebla, Mexico}

\author{X.~Wang}
\affiliation{Department of Physics, Michigan Technological University, Houghton, MI, USA}

\author{Z.~Wang}
\affiliation{Department of Physics, University of Maryland, College Park, MD, USA}

\author{I.J.~Watson}
\affiliation{University of Seoul, Seoul, Rep. of Korea}

\author{H.~Wu}
\affiliation{Department of Physics, Wisconsin IceCube Particle Astrophysics Center, University of Wisconsin-Madison, Madison, WI, USA}
\affiliation{\href{mailto:hwu298@wisc.edu}{hwu298@wisc.edu}}

\author{S.~Yu}
\affiliation{Department of Physics, Pennsylvania State University, University Park, PA, USA}

\author{C.~de León}
\affiliation{Universidad Michoacana de San Nicolás de Hidalgo, Morelia, Mexico}
 
\author{the HAWC Collaboration}

\enlargethispage{4\baselineskip}
\begin{abstract}
TeV halos are extended very-high-energy (VHE; 0.1-100~TeV) gamma-ray emission around middle-aged pulsars. So far they have only been found around isolated pulsars, but it has been suggested that they may also be powered by millisecond pulsars (MSPs). We searched for VHE gamma-ray emission from MSPs reported by radio and GeV gamma-ray observatories in 2565 days of data from the High Altitude Water Cherenkov (HAWC) Observatory. We found no significant emission from individual pulsars. By combining the likelihood profiles of all MSPs accessible to HAWC, our analysis  suggests that the excess emission around the MSP population is consistent with a background. Our result suggests that MSPs are not as efficient as isolated pulsars in producing TeV halos. This finding has strong implications on the physics interpretation of the Galactic Center GeV excess and high-latitude Galactic diffuse emission.

\end{abstract}
\maketitle

{\it Introduction.--} TeV halos (also referred to as pulsar halos) are a new category of Galactic gamma-ray sources. They have been identified around middle-aged, isolated pulsars with extensions that are significantly larger than the pulsar wind nebulae (PWNe) \cite{gemingahawc17, 3HWC, lhasso21, HAWC:2023jsq}. The extension may be explained by relativistic electrons up-scattering the Cosmic Microwave Background (CMB) in a region with a diffusion coefficient of 1-2 orders of magnitude lower than that of the average of the interstellar medium (ISM) \cite{Linden17, DiMauro2020}. The formation mechanism of TeV halos remains largely unknown \cite{Lopez-Coto:2022igd}.

While all of the identified TeV halos are located around isolated pulsars, it has been postulated that millisecond pulsars (MSPs) may also accelerate electrons to relevant energies \cite{Harding:2021yuv}. An MSP is characterized by extraordinarily short periods on the order of milliseconds. It is typically an old, rapidly rotating neutron star with a history of mass transfer from a companion star in a close binary system. MSPs are bright emitters of pulsed GeV gamma rays \cite{smith_third_2023}. Models have suggested that electrons may be accelerated in the magnetosphere \cite{Harding:2021gdt} or intrabinary shocks \cite{vanderMerwe:2020oun} further to tens of TeV energies. 

Previous works identified excess emission around several MSPs using the public HAWC data and suggested that the population may also power TeV halos at a $4\sigma$ significance level \cite{hooper_millisecond_2018, hooper_evidence_2022}. The analyses concluded that MSPs produce very-high-energy (VHE; 0.1-100~TeV) gamma-ray emission with an efficiency similar to that observed from the Geminga TeV halo, $\eta_{\rm MSP} = (0.39 - 1.08) \times \eta_{\rm Geminga}$ \cite{gemingahawc17}, where the efficiency $\eta$ is defined as the ratio of the high-energy gamma ray flux to the spin-down power of a pulsar.  

TeV halos have been invoked to explain the diffuse gamma-ray emission of the Galactic Plane. 
It has been noticed that the Galactic diffuse emission (GDE) at a few TeV measured by Milagro \cite{Milagro:2005xqq,PhysRevLett.120.121101} and above 10~TeV  measured by the Large High Altitude Air Shower Observatory (LHAASO) \cite{2023PhRvL.131o1001C} and HAWC \cite{HAWC:2023wdq} is higher than the conventional diffuse emission models describing Galactic cosmic rays interacting with the ISM gas. The ``excess" emission may be attributed to unresolved sources \cite{2022ApJ...928...19V, 2023NatAs.tmp..248F}. In particular, unresolved TeV halos have been used to explain the GDE at a few TeV \cite{PhysRevLett.120.121101, 2023arXiv230712363Y} and above 100~TeV \cite{2023arXiv230600051D}. Though above $\sim 30$~TeV, the flux of high-energy neutrinos from the Galactic Plane measured by the IceCube Observatory \cite{IceCube:2023ame} is comparable to the sum of the flux of non-pulsar sources and the diffuse emission observed by LHAASO. This suggests that the GDE above $\sim 30$~TeV is likely dominated by hadronic interactions instead of leptonic sources such as TeV halos \cite{2023ApJ...957L...6F}. Nontheless, unresolved TeV halos remain a leading interpretation of the GDE measurement at a few TeV. Understanding the TeV halo population around MSPs is crucial to the development of a complete model of the GDE.

Understanding the TeV halos powered by MSPs also has strong implications on the origin of the gamma-ray excess from the Galactic Center. An excess of GeV-scale gamma ray emission has been identified from the direction of the inner Galaxy \cite{Hooper:2010mq,Calore:2014xka,Fermi-LAT:2017opo}. It may be plausibly explained by either the annihilating dark matter particles \cite{cirelli_dark_2024} or astrophysical reasons, specifically, a group of unresolved MSPs \cite{Abazajian:2012pn,Bartels:2015aea}. \citep{hooper_millisecond_2018, hooper_evidence_2022} show that if MSPs generate TeV halos with Geminga-like efficiency, then the MSPs contributing to the GeV-scale excess should also produce TeV halo emission via inverse Compton radiation. The pulsar halo counterparts provide a way to constrain the unresolved MSP population in the Galactic Center. 


Motivated by these implications, we study the existence of TeV halos around MSPs by searching for halo-like emission around known MSPs in the HAWC data. We investigate the source confusion in the regions around the MSPs that were previously reported to have tentative halo emission. In addition, we employ a stacking technique to compare the halo signal and a background fluctuation. Each source in the pulsar samples is fitted individually using the spectral and spatial model described in the Methods section. Their log-likelihood profiles are added to constrain the common efficiency $\eta$ for each power scenario and energy bin. We find no evidence of TeV halos around the MSPs  in the northern sky observable by HAWC and report that the halo emission efficiency of the MSPs is lower than that of isolated pulsars.

{\it Methods.--} The HAWC Gamma-Ray Observatory comprises 300 water Cherenkov detectors at an elevation of 4100 meters situated at a latitude of $19^\circ$ North inside the Pico de Orizaba National Park near Puebla, Mexico. The detector possesses an instantaneous field of view that encompasses 15\% of the celestial sphere and surveys approximately two-thirds of the sky during each 24-hour period. 
The data are divided in energy bins according to the fraction of photo-multipliers tubes that are triggered by each shower event (see details in \citep{hawc2017}). In this work, the results are obtained using 2565 days of observations, spanning in energy from 300 GeV to 100 TeV.

\begin{figure}[t]
\includegraphics[width = 0.5\textwidth]{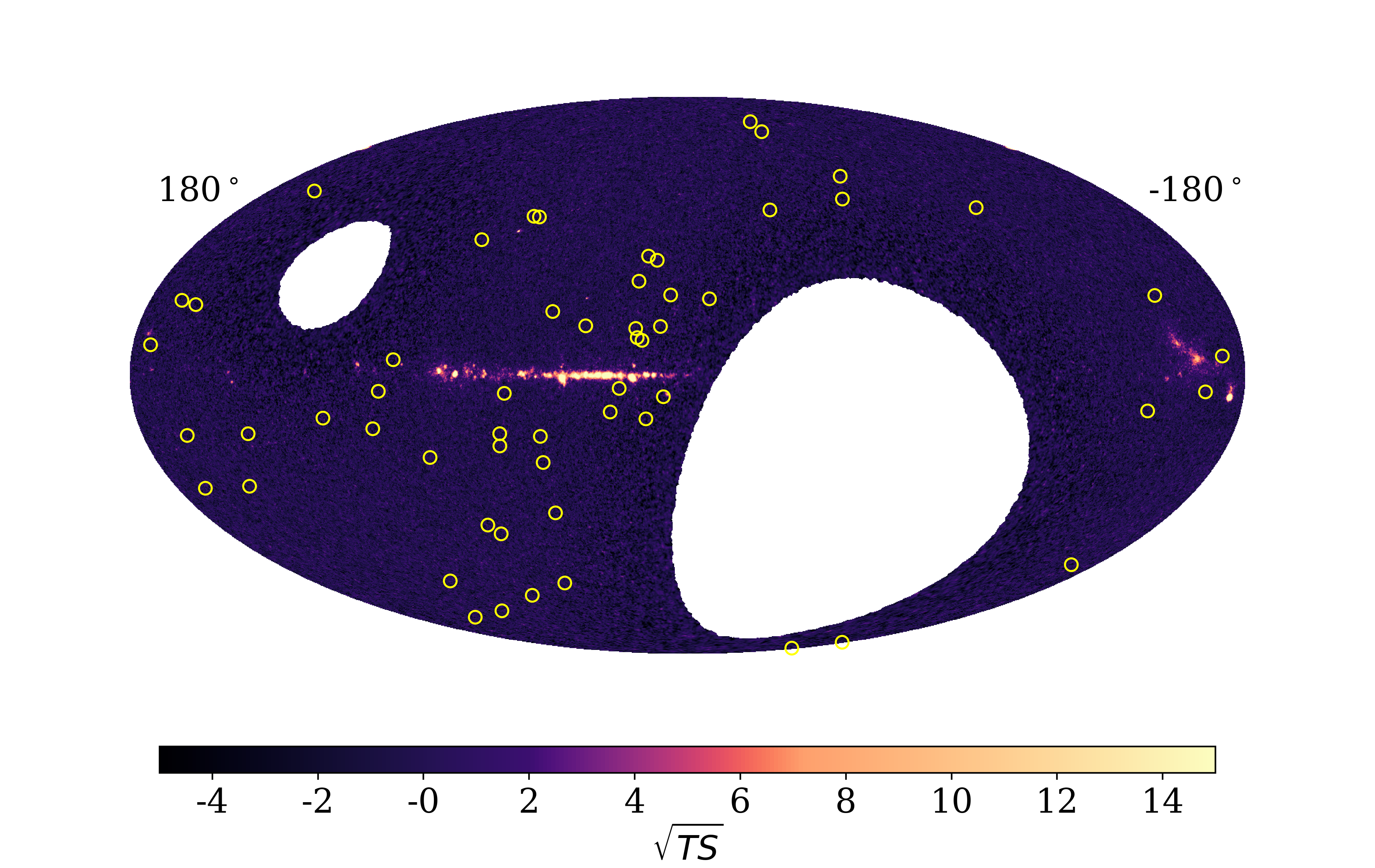}
\caption{\label{fig:gam} The HAWC significance map computed using {\it Pass 5.1} data in the Galactic coordinate. The MSPs studied our analysis are marked as yellow circles. The color bar indicates the test statistic of a point source at a given pixel assuming a power-law spectrum with an index of 2.7.}
\end{figure}

We obtain an MSP list from the Australia Telescope National Facility (ATNF) Catalog, the {\it Fermi}-LAT Pulsar Catalog (3PC) \citep{smith_third_2023}, the West Virginia University catalog (WVU) \citep{WVU_cat}, and the LOFAR Tied-Array All-Sky Survey (LOTAAS) \citep{tan_lofar_2020}. We apply the following criteria to select a set of source samples: 1) observable by HAWC ($-26^{\circ} < \delta < 64^{\circ}$), 2) away from the Galactic plane ($|b| > 3^\circ$), 3) above the HAWC sensitivity,  and 4) at least $2^\circ$ from any known sources in TeVCat \citep{TeVCat} and the 3rd HAWC catalog \citep{3HWC}. In Criterion 3), we cut off the $L_{\rm sd} / (4\pi d^2)$ at $4.12\times 10^{32} {\rm erg}\ {\rm s}^{-1} {\rm kpc}^{-2}$, the minimum detectable by HAWC for a 1\% efficiency $L_{\rm sd}$ to TeV emission at the most favorable conditions.

Criterion 4) is motivated by the fact that excess emission may be contaminated from bright sources near MSPs. Excluding MSPs near the known sources reduces the possibility of fake detection due to source confusion. 

A total of 57 sources met the criteria, including 53 from the ATNF catalog, 37 from the 3PC (with a 34 source overlap with ATNF), 3 from the WVU catalog, and 1 from LOTAAS. Among them, 45 sources are radio-loud, 12 are radio-quiet, and 39 are gamma-ray bright. Figure~\ref{fig:gam} presents the selected MSPs over the HAWC significance map. The overview plot already shows that most MSPs do not lie in regions with excess gamma-ray emission. More details about the spin-down power and distances of the selected MSPs are presented in the Appendix.  

\begin{table*}[t]
\begin{tabular}{ccccccc}\toprule
Energy Bin & Energy Range & Pivot Energy & 95\% C.I. limit on K & TS & 95\% C.I. limit on K & TS \\
 & & & ($L_{\rm sd}/(4 \pi d^2)$ ) & ($L_{\rm sd}/(4 \pi d^2)$) & ($G_{100}$) & ($G_{100}$) \\
 & [TeV] & [TeV] & [ $\rm{keV}^{-1}\rm{s}^{-1}\rm{cm}^{-2}$] & & [$\rm{keV}^{-1}\rm{s}^{-1}\rm{cm}^{-2}$] & \\ \hline
full & 0.32-100.00 & 5.62 & $9.32\times 10^{-15}$ & $-2.51\times 10^{-2}$ & $7.65\times 10^{-15}$ & $-8.43\times 10^{-3}$ \\
0 & 0.32-1.00 & 0.56 & $9.84\times 10^{-11}$ & $-8.74\times 10^{-5}$ & $1.32\times 10^{-10}$ & 1.26 \\
1 & 1.00-3.16 & 1.78 & $2.91\times 10^{-12}$ & $4.10\times 10^{-8}$ & $2.59\times 10^{-12}$ & $4.83\times 10^{-5}$ \\
2 & 3.16-10.00 & 5.62 & $1.70\times 10^{-13}$ & $2.63\times 10^{-2}$ & $1.88\times 10^{-13}$ & 2.71 \\
3 & 10.00-31.62 & 17.78 & $6.22\times 10^{-15}$ & $-1.65\times 10^{-2}$ & $3.84\times 10^{-15}$ & $-6.52\times 10^{-4}$ \\
4 & 31.62-100.00 & 56.23 & $2.57\times 10^{-16}$ & $2.49\times 10^{-5}$ & $1.22\times 10^{-16}$ & $4.05\times 10^{-5}$ \\
\toprule\end{tabular}
\caption{Flux upper limits and test statistics in the full and individual energy bins.}
\label{tab:paras_e}
\end{table*}

\begin{figure*}[t] 
    \includegraphics[width=.8\textwidth]{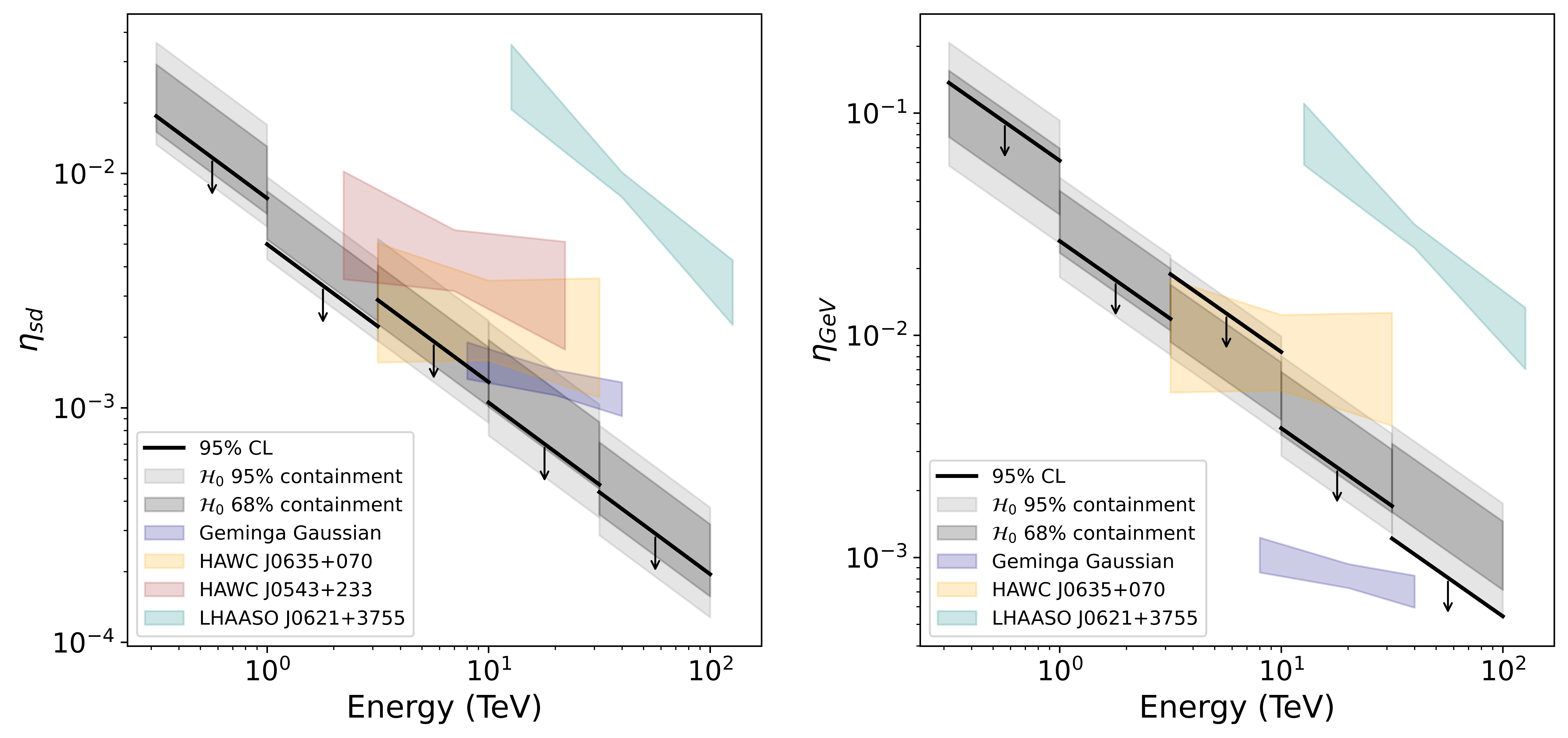}
    \caption{The efficiency refers to the fraction of TeV gamma-ray power out of the spin-down power ($L_{sd}/4\pi d^2$, left) or GeV power (right). The black solid lines with down-going arrows indicate the upper limits obtained by this analysis. The colored bands correspond to the efficiencies of observed TeV halos around middle-aged pulsars. The dark and light grey bands correspond to the 68\% and 90\% containment for expected upper limits. They are obtained by performing the same analysis but with simulated background data.}\label{fig:eta_e} 
\end{figure*}

Our data analysis is performed using {\it 3ML} (Multi-Mission Maximum Likelihood) \citep{vianello_multi-mission_2017} with the HAWC Accelerated Likelihood ({\it HAL}) plugin \citep{HAL_plugin}. We analyze the data in the energy range from 0.32 to 100.00 TeV. We use five quasi-differential energy bins with a binning scheme similar to \citep{abeysekara_measurement_2019}, described in the Appendix. Table \ref{tab:paras_e} lists the median and boundaries of the energy bins.

We model the spatial distribution of a TeV halo with a bidimensional Gaussian function. Motivated by the fact that the physical sizes of TeV halos are similar \cite{DiMauro2020}, we estimate the extension of a source by scaling from the angular size of the Geminga halo,  
\begin{equation}\label{eqn:extension}
\sigma_{\rm source} =\frac{d_\mathrm{Geminga}}{d_\mathrm{Source}} \, \sigma_\mathrm{Geminga}
\end{equation}
where $d_\mathrm{Geminga} = 250\mathrm{pc}$ and $d_\mathrm{Source}$ are the distances to the Geminga pulsar and the source, respectively, $\sigma_\mathrm{Geminga} = 2.0 ^{\circ}$ is Geminga's angular extension under a Gaussian model \citep{Linden17}.

The gamma ray flux $\Phi$ of the $i$th source can be described as a power-law:
\begin{equation}
    \Phi_i=K C_i\left(\frac{E}{E_{\mathrm{piv}}}\right)^{-p},
\end{equation}
where the pivot energy $E_{\rm piv}$ is set to be the medians of the energy bins in logarithmic scale, the spectral index $p$ is set to the mean index of the HAWC sources \cite{2HWC}, $p = 2.7$, $C_i$ is a source-dependent term, denoted as a source weight, and $K$ is a universal normalization factor for all sources which is left as a free parameter in the fit. The uncertainty of the result associated with the choice of $p$ is later studied as a systematic error.

We define a test statistic (TS) to be twice the logarithm of the likelihood ratio when fitting the data with and without the above-mentioned TeV halo model, ${\rm TS} \equiv2 \ln [ {\cal L}(\hat{K}) / {\cal L} (K = 0)]$.   

For an individual source, $K$ and $C_i$ are coupled so we can set $C_i = 1$. For a stacking analysis where we combine the likelihoods of TeV halos around a sample of sources, we consider two source weighting schemes. In the first scenario, denoted as the spin-down weighting scheme, we assume that the TeV halo flux of an MSP is proportional to the spin-down flux ($L_{\rm sd}$) of the pulsar, $C_i \propto L_{{\rm sd},i}/ (4\pi\,d_i^2)$. In the second scenario, denoted as the GeV weighting scheme, we assume that the TeV halo flux scales with the gamma-ray flux of the pulsar in 0.1--100~GeV measured by {\it Fermi}-LAT (named $G_{100}$ in 3PC), $C_i \propto G_{100}$. As the formation mechanism of TeV halos is largely unknown, these scenarios serve as general assumptions that link the TeV flux to the power of MSPs. The values of $C_i$ are normalized such that $\sum C_i = 1$ for the whole sample.

{\it Results.--} We first examine the MSPs with tentative TeV halo detection reported by previous works \cite{hooper_millisecond_2018, hooper_evidence_2022}. We find no significant emission at the positions of target MSPs (see Appendix for the fitting results of individual sources).
Our study suggests that the previous tentative detections reported by \cite{hooper_millisecond_2018, hooper_evidence_2022} are consistent with statistical fluctuations.

To study whether TeV halos commonly exist around the MSP population, we combine the likelihood profiles of the source sample and compare them with a null hypothesis (the non-existence of TeV halos around MSPs). We use all 57 sources in the spin-down weighting scheme analysis and the 39 gamma-ray bright sources in the GeV weighting scheme analysis. Table~\ref{tab:paras_e} summarizes the results of the stacking searches. In all weighting schemes and energy ranges, we find that the stacked TS is consistent with which is expected from background fluctuations.

We compare the TeV halo emission efficiency of the MSPs to that of the observed halos. The efficiency is defined as the ratio of the differential flux in a given energy bin with respect to the spin-down or the GeV flux of the pulsar:
\begin{equation}
\eta_{\rm sd} \equiv \frac{E^2 \Phi_i}{L_{{\rm sd},i} / (4 \pi d^2)}; \quad \eta_{\rm GeV} \equiv \frac{E^2 \Phi_i}{G_{100,i}}. 
\label{eqn:eta}
\end{equation}
Note that $\eta$ is assumed to be identical for all pulsars in a stacking analysis. Figure~\ref{fig:eta_e} presents the halo emission efficiency in the spin-down and GeV gamma-ray flux scenarios. For comparison, we show the efficiency for the detected TeV halos, including Geminga \citep{gemingahawc17}, HAWC J0635+070 \citep{brisbois_hawc_2018}, HAWC J0543+233 \citep{riviere_hawc_2017} and LHAASO J0621+3755 \citep{aharonian_extended_2021}. 
Our limits suggest that the TeV halo emission efficiency of MSPs is lower than the observed TeV halos, especially for above 10~TeV. 

The absolute flux sensitivity error is at the level of $\sim 30\%$ for the instrument based on our understanding of the absolute detector performance and temporal variability. The factors that contribute to the systematic uncertainties are explained in the Appendix. 

{\it Discussion. --} With the latest HAWC data, we find that the tentative TeV halo emission around the MSPs reported by previous works is consistent with background fluctuations. By adding the likelihoods (log likelihood vs. efficiency) functions of halo-like emission around an MSP population under the assumption that the TeV flux is proportional to the spin-down or GeV flux of the pulsar, we conclude that MSPs are not as efficient as isolated pulsars in emitting TeV halos. In addition, we find that the point-like emission from the MSP population is also consistent with the background, though the upper limits on individual bright MSPs are approaching the theoretical models, as further described in the Appendix.

The non-detection of TeV halos around MSPs may be caused by one or a combination of  the following reasons. First, VHE electron acceleration by MSPs, which could happen in the pulsar magnetosphere or the intrabinary shocks, may not be as efficient as that by middle-aged pulsars and their nebulae. Second, the extensions of the observed TeV halos indicate strong confinement of electrons in the vicinity of the pulsars. The origin of this confinement is still poorly understood. Some models attribute the slow diffusion to a turbulence driven by the central source \cite{2018PhRvD..98f3017E, 2019MNRAS.488.4074F, 2022PhRvD.105l3008M}.  With intrabinary activities in regions much smaller than PWNe, MSPs could be incapable of producing such slow diffusion regions.   

While GeV emission of MSPs could be as bright as that of middle-aged pulsars, our study suggests that the TeV halos of MSPs should not significantly contribute to the GDE at TeV or higher energies. The GDE flux at a few TeV may still be explained by unresolved TeV halos by isolated, middle-aged pulsars \cite{eckner_detecting_2022}, which are mainly distributed along the Galactic Plane.

The non-detection of MSP TeV halos observable by HAWC breaks down the degeneracy of {\it Fermi}-LAT excess and VHE emission by MSPs in the Galactic Center region. The lower halo production efficiency found by our work suggests that MSPs may produce GeV gamma-ray emission without significant VHE and radio emission. The constraints posed by \citep{hooper_millisecond_2018, hooper_evidence_2022} by comparing the TeV halo flux of MSPs with the spin-down power needed to explain the GeV excess and the H.E.S.S. observation of the Galactic Center region is significantly weakened. Our results support that MSPs remain one of the plausible explanations to the Galactic Center gamma-ray excess.

\vspace{2em}

\begin{acknowledgments}
We acknowledge the support from: the US National Science Foundation (NSF); the US Department of Energy Office of High-Energy Physics; the Laboratory Directed Research and Development (LDRD) program of Los Alamos National Laboratory; Consejo Nacional de Ciencia y Tecnolog\'{i}a (CONACyT), M\'{e}xico, grants LNC-2023-117, 271051, 232656, 260378, 179588, 254964, 258865, 243290, 132197, A1-S-46288, A1-S-22784, CF-2023-I-645, c\'{a}tedras 873, 1563, 341, 323, Red HAWC, M\'{e}xico; DGAPA-UNAM grants IG101323, IN111716-3, IN111419, IA102019, IN106521, IN114924, IN110521 , IN102223; VIEP-BUAP; Programa Integral de Fortalecimiento Institucional (PIFI) 2012, 2013, Programa de Fortalecimiento de la Calidad Educativa (PROFOCIE) 2014, 2015; the University of Wisconsin Alumni Research Foundation; the Institute of Geophysics, Planetary Physics, and Signatures at Los Alamos National Laboratory; Polish Science Centre grant, DEC-2017/27/B/ST9/02272; Coordinaci\'{o}n de la Investigaci\'{o}n Cient\'{i}fica de la Universidad Michoacana; Royal Society - Newton Advanced Fellowship 180385; Generalitat Valenciana, grant CIDEGENT/2018/034; The Program Management Unit for Human Resources \& Institutional Development, Research and Innovation, NXPO (grant number B16F630069); Coordinaci\'{o}n General Acad\'{e}mica e Innovaci\'{o}n (CGAI-UdeG), PRODEP-SEP UDG-CA-499; Institute of Cosmic Ray Research (ICRR), University of Tokyo. H.F. acknowledges support by NASA under award number 80GSFC21M0002. We also acknowledge the significant contributions over many years of Stefan Westerhoff, Gaurang Yodh and Arnulfo Zepeda Dom\'inguez, all deceased members of the HAWC collaboration. Thanks to Scott Delay, Luciano D\'{i}az and Eduardo Murrieta for technical support.
\end{acknowledgments}

\newcommand {\apjl}     {ApJ Letters}
\newcommand {\mnras}    {MNRAS}
\bibliographystyle{apsrev4-1}
\bibliography{references}

\clearpage 
\appendix
\section{Appendices}

\subsection{Properties of the source sample}\label{app:sourceSample}

\begin{figure}[t]
    \centering
    \includegraphics[width=.45\textwidth]{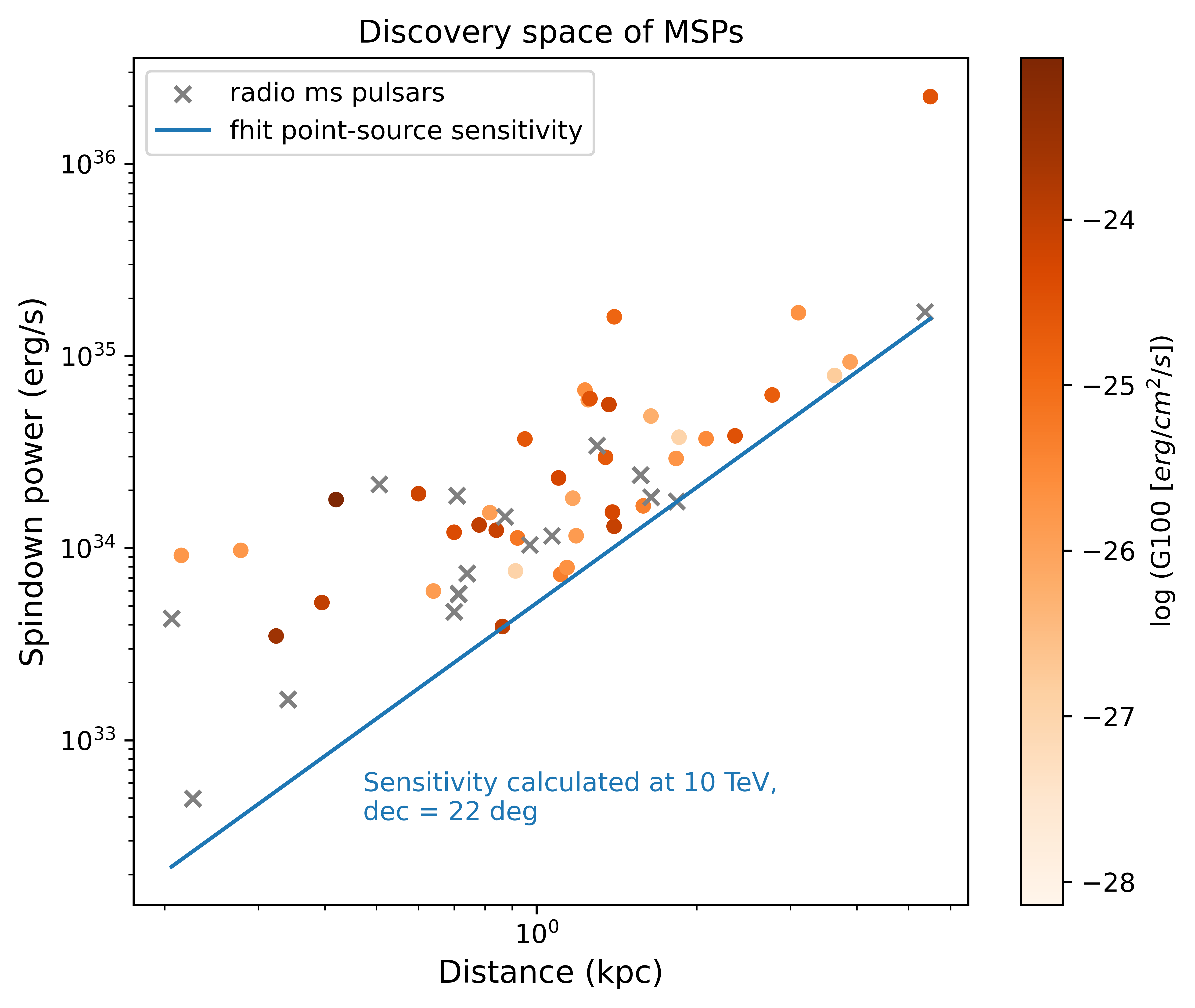}
    \caption{\label{fig:dis_sp} The spin-down power and distances of the MSPs that satisfy our selection criteria. The blue curve indicates the sensitivity of HAWC for a point-like source at $\delta = 22^\circ$ integrated from 0.316 TeV to 100 TeV. A total of 57 sources are shown, with the color indicating G100. }
\end{figure}

Figure~\ref{fig:dis_sp} presents the spin-down power and distances of the MSPs in our sample. The colored markers indicate the gamma-ray-bright MSPs and the cross markers indicate the gamma-ray-dim MSPs detected in the radio band. 

The HAWC point-source flux sensitivity is defined as the flux normalization required to have a 50\% probability of detecting a source at the $5\sigma$ level \cite{albert_performance_2024}. For a point source at $\delta = 22^\circ$, the detector's sensitivity is $\Phi = 2.7\times10^{-15} \left(E / 10\, {\mathrm {TeV}}\right)^{-2.63}\, {\mathrm{TeV^{-1}\,cm^{-2}\, s^{-1}}}$. By converting the differential gamma-ray flux to the pulsar spin-down flux using an efficiency $\eta_{\rm sd}= 1\%$ at the pivot energy $E_{\rm piv} = 10$~TeV, we obtain a selection criterion on the pulsar spin-down flux, $L_{\rm sd} / (4\pi d^2) \geq E_{\rm piv}^2 \Phi(E_{\rm piv}) / \eta  = 4.12\times 10^{32} {\rm erg}\ {\rm s}^{-1} {\rm kpc}^{-2}$. As the detector sensitivity would be worse at other declinations and for extended sources, also because that the efficiency of the known TeV halos is lower than 1\%, pulsars with spin-down flux below this conservative selection criterion cannot be detected by HAWC. 
\subsection{MSPs with tentative detection in previous works}\label{app:comp}

A total of 37 MSPs are considered in \citet{hooper_evidence_2022}, which are selected based on $-20^{\circ} < \delta < 50^{\circ}$ and $\dot E/d^2 > 5\times 10^{33}\, \mathrm{erg}\,\mathrm{kpc}^{-2} \,{\rm s}^{-1}$. Among them, four sources are found with $({\rm TS})^{1/2} \geq 2.55$ using the public HAWC software, which includes data from HAWC's fourth pass.

\begin{figure}[t]
\includegraphics[width=0.45\textwidth]{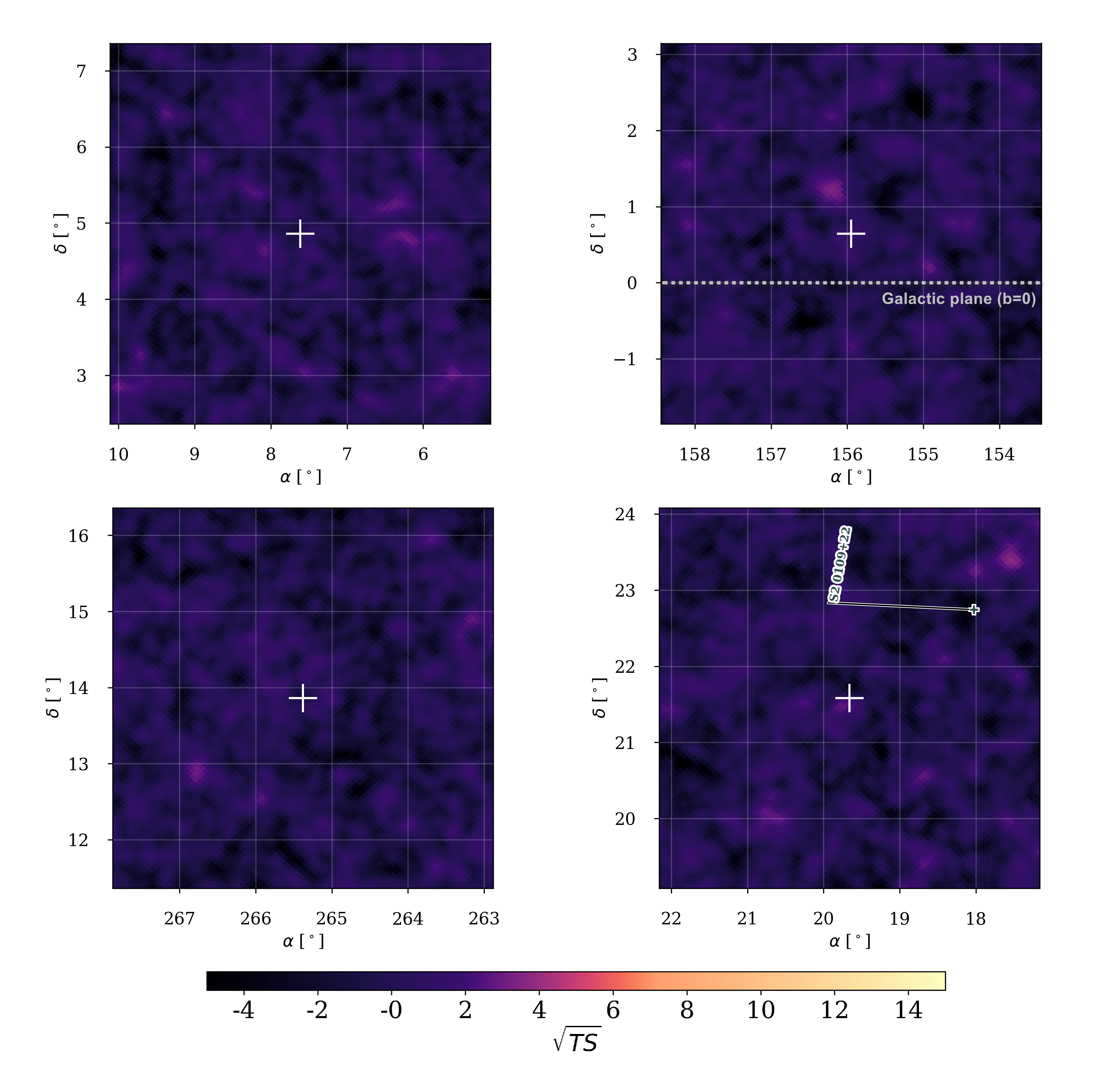}
\caption{Locations of 4 pulsars with $\sqrt{TS} > 2.55$ in Hooper and Linden's sample. The sources at the center are PSRJ0030+0451 (top-left), PSRJ1023+0038 (top-right), PSRJ1741+1351 (bottom-left), and PSRJ1939+2134 (bottom-right), respectively. Nearby sources from the TeVCat are labeled.}
\label{fig:TSmap}
\end{figure}

We use data from HAWC's fifth pass \citep{albert_performance_2024} through the data with updated reconstructed algorithms to inspect these sources. Main improvements from the fourth pass to the fifth pass include an increase of data from 1523 days to 2565 days, new methodologies to identify and remove the background in lower energy analyses, enhancement of the directional reconstruction for high-energy showers, and the gamma/hadron separation efficiency. Table \ref{tab:HLsources} lists the $({\rm TS})^{1/2}$ of each of the MSPs when fitting a point-like source with a power-law spectrum $dN/dE\propto E^{-2.7}$. All sources have $({\rm TS})^{1/2} < 2$. Figure~\ref{fig:TSmap} presents the HAWC significance maps of the four tentative sources found by \citet{hooper_evidence_2022}. Gamma-ray emission in the regions is consistent with background fluctuations with more data and better event reconstruction. The detection in \citet{hooper_evidence_2022} is most likely a result of detecting stray galactic plane emission from sources near the galactic plane. In our work, this excess emission is absent due to the exclusion of the Galactic plane during source selection.

\begin{table}
\begin{tabular}{ccccc}
\toprule
Name & RA & Dec & $TS^{1/2}$ & $TS^{1/2}$\\
 & [$^\circ$] & [$^\circ$] & (this work) & (HL22)\\\hline
PSRJ0023+0923 & 5.82 & 9.39 & 1.34 & 1.33 \\
PSRJ0030+0451 & 7.61 & 4.86 & 0.94 & 2.55 \\
PSRJ0034-0534 & 8.59 & -5.58 & 1.92 & 0.10 \\
PSRJ0218+4232 & 34.53 & 42.54 & 1.52 & 1.56 \\
PSRJ0337+1715 & 54.43 & 17.25 & 1.54 & 0.25 \\
PSRJ0557+1550 & 89.38 & 15.84 & -0.15 & 0.14 \\
PSRJ0605+3757 & 91.27 & 37.96 & $-8.20\times 10^{-2}$ & -1.02 \\
PSRJ0613-0200 & 93.43 & -2.01 & 1.73 & $6.00\times 10^{-2}$ \\
PSRJ0621+2514 & 95.30 & 25.23 & -0.10 & 1.62 \\
PSRJ0751+1807 & 117.79 & 18.13 & $-1.44\times 10^{-2}$ & -2.09 \\
PSRJ1023+0038 & 155.95 & 0.64 & 0.83 & 2.56 \\
PSRJ1231-1411 & 187.80 & -14.20 & $-4.99\times 10^{-2}$ & $-2.00\times 10^{-2}$ \\
PSRJ1300+1240 & 13.00 & 12.68 & -0.14 & -0.59 \\
PSRJ1400-1431 & 210.15 & -14.53 & 0.12 & -1.04 \\
PSRJ1622-0315 & 245.75 & -3.26 & 0.88 & -0.51 \\
PSRJ1630+3734 & 247.65 & 37.58 & 1.40 & -0.59 \\
PSRJ1643-1224 & 250.91 & -12.42 & $-4.29\times 10^{-2}$ & 0.66 \\
PSRJ1710+4923 & 257.52 & 49.39 & 0.66 & -0.62 \\
PSRJ1719-1438 & 259.79 & -14.63 & $-4.61\times 10^{-2}$ & -0.56 \\
PSRJ1737-0811 & 264.45 & -8.19 & $-9.14\times 10^{-2}$ & -0.16 \\
PSRJ1741+1351 & 265.38 & 13.86 & $-7.93\times 10^{-2}$ & 2.64 \\
PSRJ1744-1134 & 266.12 & -11.58 & $-6.74\times 10^{-2}$ & -0.95 \\
PSRJ1745-0952 & 266.29 & -9.88 & 1.23 & -1.97 \\
PSRJ1843-1113 & 280.92 & -11.23 & 1.76 & 0.15 \\
PSRJ1911-1114 & 287.96 & -11.24 & $-6.82\times 10^{-2}$ & $-2.00\times 10^{-2}$ \\
PSRJ1921+1929 & 290.35 & 19.49 & $3.38\times 10^{-3}$ & 0.62 \\
PSRJ1939+2134 & 19.66 & 21.58 & 0.83 & 3.34 \\
PSRJ1959+2048 & 19.99 & 20.80 & 0.95 & 2.12 \\
PSRJ2017+0603 & 304.34 & 6.05 & 0.13 & -0.37 \\
PSRJ2042+0246 & 310.55 & 2.77 & 0.40 & -0.67 \\
PSRJ2043+1711 & 310.84 & 17.19 & $-8.46\times 10^{-3}$ & -0.72 \\
PSRJ2214+3000 & 333.66 & 30.01 & 1.98 & 0.33 \\
PSRJ2234+0611 & 338.60 & 6.19 & -0.11 & -0.23 \\
PSRJ2234+0944 & 338.70 & 9.74 & 1.34 & 0.80 \\
PSRJ2256-1024 & 344.23 & -10.41 & 1.21 & 0.45 \\
PSRJ2302+4442 & 345.70 & 44.71 & $-9.28\times 10^{-2}$ & $-1.00\times 10^{-2}$ \\
PSRJ2339-0533 & 354.91 & -5.55 & 0.63 & -0.35 \\
\toprule\end{tabular}
\caption{Hooper/Linden's MSP list. ${\rm TS}^{1/2}$ in this work is obtained by fitting a point-like source with a power-law spectrum $dN/dE\propto E^{-2.7}$ to the HAWC pass~5.1 data. Negative signs of ${\rm TS}^{1/2}$ stand for negative TS. ${\rm TS}^{1/2}$ (HL22) is the square root TS obtained from \citet{hooper_evidence_2022}.}
\label{tab:HLsources}
\end{table}

\subsection{Distribution of the Stacked TS }\label{app:TS_BG}
\begin{figure}[t!]
    \includegraphics[width=0.4\textwidth]{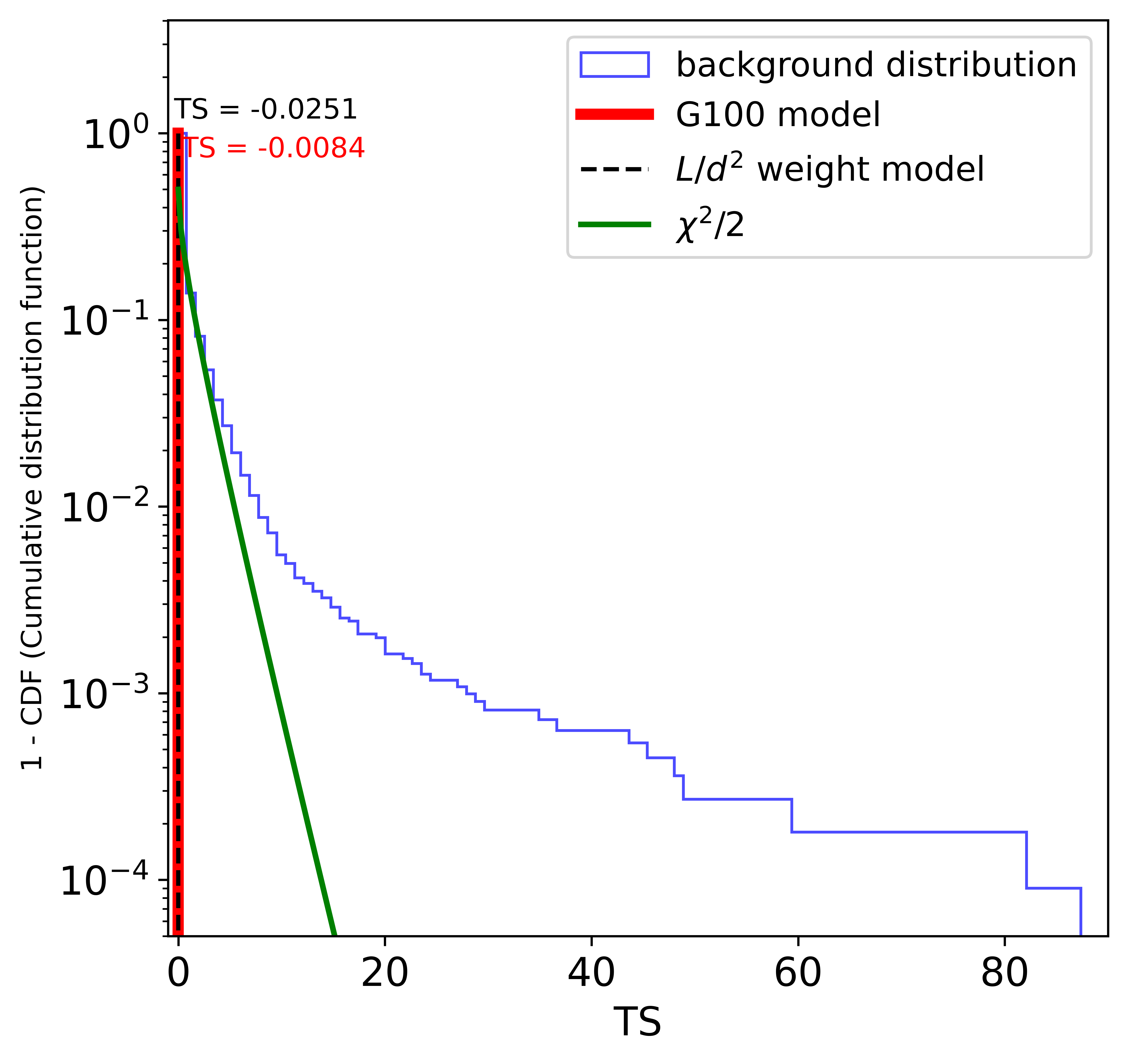}
    \caption{TS of full energy range fits in our extended source assumption analysis, compared with background TS distribution and $\chi^2/2$. The background distribution is composed of a total of 11068 trials. 10 random locations with a hypothetical extension of $0.468^\circ$, which is the mean extension of our source sample, are stacked in each trial.}
    \label{fig:back-ext}
\end{figure}

We use Monte Carlo simulations to study the TS distribution of the null hypothesis. We generate random positions in the sky and use the same selection criteria as for the real sources to form a fake source sample. Specifically, we require that a fake source is in the HAWC's field-of-view and located at least $2^\circ$ away from TeVCat sources. We assume that all fake sources have the same extension of $0.4^\circ$, which is close to the median extension $0.41$ of the actual source sample. The spatial template is assumed to be Gaussian, the same as for the real sample.

In the simulation, we first generate a fake source sample composed of 10 random positions that satisfy the source selection criteria, fit each source in the sample, and stack the likelihoods to find a stacked TS. We then repeat the process for 3093 times. The distribution of the resulted TS is shown as the blue curve in Figure~\ref{fig:back-ext}.  The background distribution departs from a $\chi^2/2$ distribution expected from Poisson statistics. The tail of the distribution is caused by the random sources located near the real sources that are more extended than $\sim 2^\circ$. Stacking fake source samples that are as large as our actual sample is computationally expensive but we have verified that the TS distribution is insensitive on source number up to a sample size of 40.

Based on the background distribution, we identify the TS value that corresponds to the p-value of $2\sigma$ in a two-sided Gaussian distribution as ${\rm TS}_{2\sigma} = 3.99$. 

The stacked TS obtained from the real source sample under the spin-down and GeV weighting schemes are presented as the dashed black curve and solid red curve, respectively. Both are consistent with the background TS distribution.

\subsection{Stacking analysis of point-like emission from MSPs}\label{app:point-like}

\begin{figure*}[t] 
    \includegraphics[width=.8\textwidth]{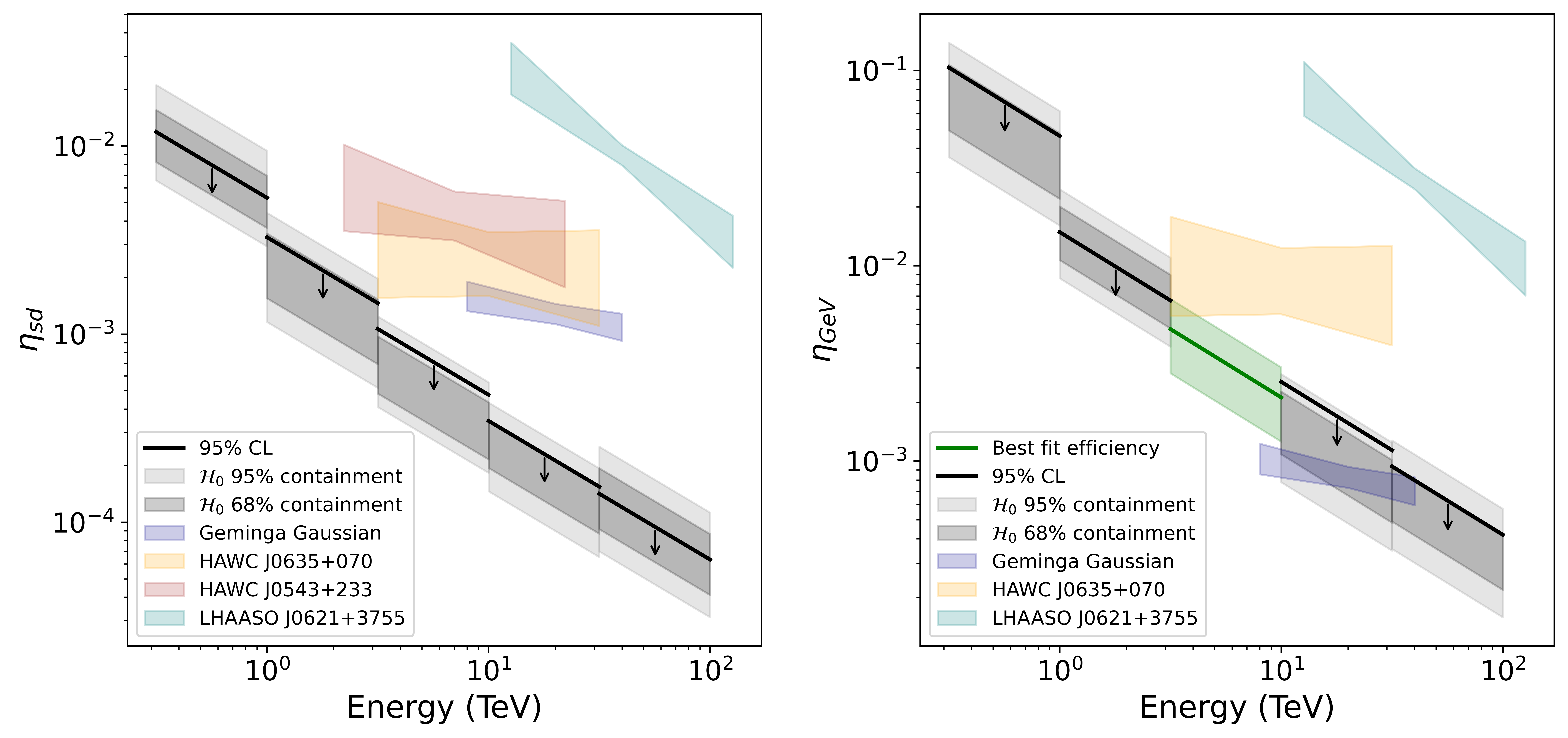}
    \caption{The point-source model efficiency of the spin-down power ($L_{sd}/4\pi d^2$, left) or GeV power (right). Detection with $TS > 4$ are marked with colored uncertainty bands while non-detection upper limits are marked with grey Brazilian bands. }\label{fig:eta_p}
\end{figure*}

In addition to the extended source model presented in the main text, we also perform a stacking analysis of the MSP sample using a point-like emission model. We present the fitting results in table~\ref{tab:paras_p} and the efficiency as function of energy in Fig.~\ref{fig:eta_p}. Our analysis yields marginal detection, with ${\rm TS} > 4$ or equivalently at $2\sigma$ level, in the full energy range as well as bin 2 and 3. 

\begin{table*}[t]
\begin{tabular}{ccccccc}\toprule
Energy Bin & Energy Range & Pivot Energy & 95\% C.I. limit on K & TS & 95\% C.I. limit on K & TS \\
 & & & ($L_{\rm sd}/(4 \pi d^2)$ ) & ($L_{\rm sd}/(4 \pi d^2)$) & ($G_{100}$) & ($G_{100}$) \\
 & [TeV] & [TeV] & [$\rm{keV}^{-1}\rm{s}^{-1}\rm{cm}^{-2}$] & & [$\rm{keV}^{-1}\rm{s}^{-1}\rm{cm}^{-2}$] & \\ \hline
full & 0.32-100.00 & 5.62 & $4.9\times 10^{-15}$ & 2.43 & $6.3\times 10^{-15}$ & 5.04 \\
0 & 0.32-1.00 & 0.56 & $7.4\times 10^{-11}$ & 0.30 & $8.9\times 10^{-11}$ & 0.74 \\
1 & 1.00-3.16 & 1.78 & $2.1\times 10^{-12}$ & 1.78 & $1.7\times 10^{-12}$ & 0.28 \\
2 & 3.16-10.00 & 5.62 & $6.6\times 10^{-14}$ & 2.37 & $9.7\times 10^{-14}$ & 6.57 \\
3 & 10.00-31.62 & 17.78 & $2.2\times 10^{-15}$ & 1.06 & $3.1\times 10^{-15}$ & 4.18 \\
4 & 31.62-100.00 & 56.23 & $8.4\times 10^{-17}$ & 0.43 & $8.9\times 10^{-17}$ & 0.65 \\
\toprule\end{tabular}
\caption{Flux upper limits and test statistics in the full and individual energy bins under the point-source model.}
\label{tab:paras_p}
\end{table*}

The fits yield TS values close to 0 for most MSPs except the following sources. PSR~J$1911+1347$ is the most significant source with ${\rm TS}=21.1$. In addition, PSR~B$1855+09$ (${\rm TS}=6.5$) and PSR~J$0030+0451$ (${\rm TS}=4.5$) are the only ones that result in $4<{\rm TS}<9$. 

As the best fit TS of the two weight models over the full energy range is 2.43 and 5.04, which indicates that the tentative signal is well below $5\sigma$, we conclude that the point-like emission of the MSPs in the source sample is consistent with the null hypothesis. 

\subsection{Systematic Uncertainties}\label{app:syst}
\begin{table*}[h]
\centering
\begin{tabular}{c|cccc|cccc}
\toprule
\multirow{2}{*}{EBIN} & \multicolumn{4}{c|}{$L_{\rm sd}/(4 \pi d^2)$} & \multicolumn{4}{c}{$G_{100}$} \\
& Index & Response & Extension & Total & Index & Response &  Extension & Total \\ \hline
1 & $^{+2.64\%}_{-5.28\%}$ & $^{+7.72\%}_{-19.00\%}$ & $\pm22.66\%$ & $^{+24.09\%}_{-30.04\%}$ & $^{+0.76\%}_{-18.94\%}$ & $^{+8.33\%}_{-20.45\%}$ & $\pm7.58\%$ & $^{+11.29\%}_{-28.89\%}$ \\
2 & $^{+1.37\%}_{-5.84\%}$ & $^{+9.97\%}_{-10.31\%}$ & $\pm13.06\%$ & $^{+16.48\%}_{-17.63\%}$ & $^{+2.70\%}_{-7.34\%}$ & $^{+16.22\%}_{-13.90\%}$ & $\pm7.72\%$ & $^{+18.16\%}_{-17.51\%}$ \\
3 & $^{+1.18\%}_{-7.06\%}$ & $^{+0.59\%}_{-20.59\%}$ & $\pm41.18\%$ & $^{+41.20\%}_{-46.57\%}$ & $^{+1.06\%}_{-2.13\%}$ & $^{+2.66\%}_{-21.81\%}$ & $\pm9.04\%$ & $^{+9.49\%}_{-23.70\%}$ \\
4 & $^{+6.11\%}_{-2.57\%}$ & $^{+1.93\%}_{-22.03\%}$ & $\pm6.27\%$ & $^{+8.96\%}_{-23.04\%}$ & $^{+5.99\%}_{-3.65\%}$ & $^{+1.04\%}_{-26.82\%}$ & $\pm11.72\%$ & $^{+13.20\%}_{-29.50\%}$ \\
5 & $^{+1.56\%}_{-4.28\%}$ & $^{+3.11\%}_{-21.40\%}$ & $\pm45.91\%$ & $^{+46.05\%}_{-50.84\%}$ & $^{+0.82\%}_{-7.38\%}$ & $^{+8.20\%}_{-21.39\%}$ & $\pm17.21\%$ & $^{+19.08\%}_{-28.43\%}$ \\
\toprule\end{tabular}
\caption{Extended model systematics. The first 2 columns of systematics using each weight model are obtained by varying the spectral index and detector response file, or using the same extension (the mean extension $0.468^\circ$) for the whole sample}. The systematics are obtained as the difference between the base case and the maximum (above) or minimum (below) fitted result obtained from the corresponding parameter variations. The total systematics is a quadrature sum of all systematics groups.
\label{tab:syst-ext}
\end{table*}

Several factors contribute to the systematic uncertainty of our study: 1) detector configurations, 2) model uncertainties including the spectral index and source extension.

Table \ref{tab:syst-ext} summarizes the key systematic uncertainties of our study. We study two causes of systematic errors, namely, the uncertainty associated with the unknown spectral index and the detector response function. Table V presents a comprehensive overview of all sources included in the sample.

We evaluate the systematic uncertainty in five energy bins and compute the errors as the difference in the 95\% upper limits when varying the corresponding parameter. We calculate the total uncertainty as the quadrature sum of  the systematic uncertainties due to index and detector response.  

\subsection{Magnetospheric emission of individual MSPs}
\begin{figure*} 
    \includegraphics[width=.4\textwidth]{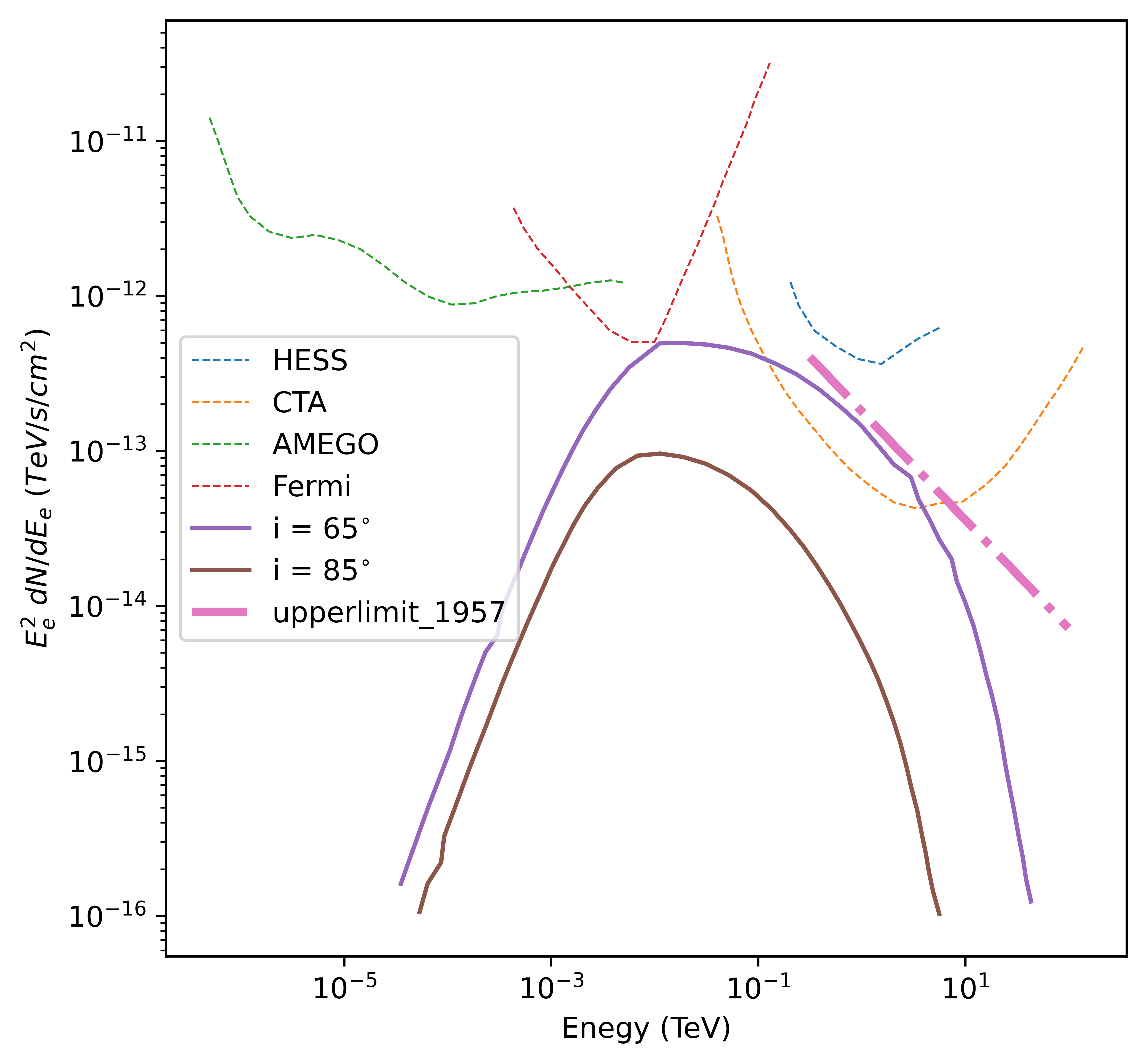}
    \centering
    \caption{Comparison of model SED plots for PSR B1957+20 depicting the effect of varying inclination $i$.}\label{fig:1957-up}
\end{figure*}

\begin{figure*} 
    \includegraphics[width=.4\textwidth]{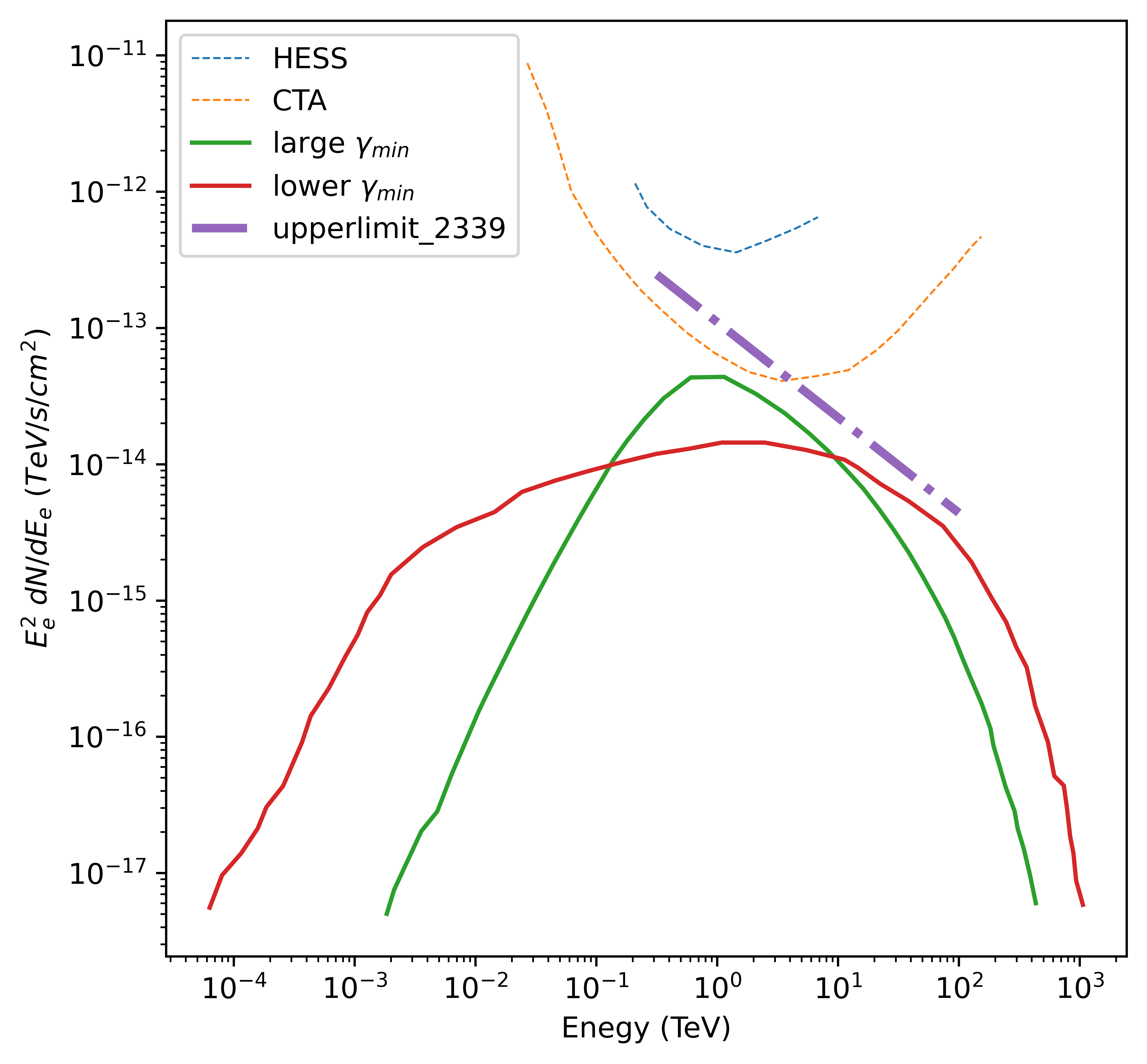}
    \centering
    \caption{Comparison of model SED plots for PSR J2339$-$0533 depicting  two different fitting scenarios: (1) the particle spectrum's low-energy cutoff (${\gamma }_{\min}$) is large so that we fit X-ray data using the intrinsic single-particle SR spectral slope of 4/3 (green line), and (2) using a lower ${\gamma}_{\min}$ so that the X-ray spectrum is matched by varying the particle spectral index p (red line).}\label{fig:2339-up}
\end{figure*}

TeV gamma-ray emission is predicted to be produced in the MSP magnetosphere. 
Figure~\ref{fig:1957-up} and Figure~\ref{fig:2339-up} compares our upper limits to the models of PSR~B1957+20 and PSR~J2339-0533 from \citet{van_der_merwe_x-ray_2020}. The HAWC limits are above the model prediction but suggest that a detection will be possible with a $25-60\%$ better sensitivity. This may be achieved with future observations of HAWC and the next-generation air shower experiments like the Cherenkov Telescope Array (CTA) \cite{eckner_detecting_2022}.

\begin{table*}
{\fontsize{10}{11}\selectfont
\begin{tabular}{|c|c|c|c|c|c|c|c|c|}\toprule
NAME & RA & Dec & EDOT & DIST & G100 & Extension & C (L/d2) & C (G100) \\\hline
PSRJ0605+3757 & 91.27 & 37.96 & $9.17\times 10^{33}$ & 0.21 & $6.44\times 10^{-12}$ & 2.33 & 0.12 & $8.45\times 10^{-3}$ \\
PSRJ1400-1431 & 210.15 & -14.53 & $9.74\times 10^{33}$ & 0.28 & $6.42\times 10^{-12}$ & 1.80 & $7.48\times 10^{-2}$ & $8.42\times 10^{-3}$ \\
PSRJ1231-1411 & 187.80 & -14.20 & $1.79\times 10^{34}$ & 0.42 & $1.01\times 10^{-10}$ & 1.19 & $6.03\times 10^{-2}$ & 0.13 \\
PSRJ1737-0811 & 264.45 & -8.19 & $4.30\times 10^{33}$ & 0.21 &  & 2.43 & $6.02\times 10^{-2}$ &  \\
PSRJ1737-0314A & 17.63 & -8.19 & $4.30\times 10^{33}$ & 0.21 &  & 2.43 & $6.02\times 10^{-2}$ &  \\
PSRJ1710+4923 & 257.52 & 49.39 & $2.15\times 10^{34}$ & 0.51 &  & 0.99 & $4.99\times 10^{-2}$ &  \\
PSRB1957+20 & 299.90 & 20.80 & $1.60\times 10^{35}$ & 1.40 & $1.57\times 10^{-11}$ & 0.36 & $4.85\times 10^{-2}$ & $2.06\times 10^{-2}$ \\
PSRB1821-24A & 276.13 & -24.87 & $2.24\times 10^{36}$ & 5.50 & $2.24\times 10^{-11}$ & $9.09\times 10^{-2}$ & $4.40\times 10^{-2}$ & $2.94\times 10^{-2}$ \\
PSRJ2214+3000 & 333.66 & 30.01 & $1.92\times 10^{34}$ & 0.60 & $3.26\times 10^{-11}$ & 0.83 & $3.17\times 10^{-2}$ & $4.28\times 10^{-2}$ \\
PSRJ1301+0833 & 195.41 & 8.57 & $6.65\times 10^{34}$ & 1.23 & $7.69\times 10^{-12}$ & 0.41 & $2.60\times 10^{-2}$ & $1.01\times 10^{-2}$ \\
PSRJ1625-0021 & 246.29 & -0.36 & $3.70\times 10^{34}$ & 0.95 & $2.07\times 10^{-11}$ & 0.53 & $2.43\times 10^{-2}$ & $2.72\times 10^{-2}$ \\
PSRJ1221-0633 & 185.35 & -6.56 & $5.92\times 10^{34}$ & 1.25 & $5.82\times 10^{-12}$ & 0.40 & $2.25\times 10^{-2}$ & $7.64\times 10^{-3}$ \\
PSRJ1843-1113 & 280.92 & -11.23 & $6.00\times 10^{34}$ & 1.26 & $2.28\times 10^{-11}$ & 0.40 & $2.24\times 10^{-2}$ & $2.99\times 10^{-2}$ \\
PSRB1257+12 & 195.01 & 12.68 & $1.88\times 10^{34}$ & 0.71 &  & 0.71 & $2.22\times 10^{-2}$ &  \\
PSRJ1744-1134 & 266.12 & -11.58 & $5.21\times 10^{33}$ & 0.40 & $3.71\times 10^{-11}$ & 1.27 & $1.98\times 10^{-2}$ & $4.87\times 10^{-2}$ \\
PSRJ0030+0451 & 7.61 & 4.86 & $3.49\times 10^{33}$ & 0.32 & $6.04\times 10^{-11}$ & 1.54 & $1.97\times 10^{-2}$ & $7.92\times 10^{-2}$ \\
PSRJ1023+0038 & 155.95 & 0.64 & $5.59\times 10^{34}$ & 1.37 & $3.24\times 10^{-11}$ & 0.37 & $1.77\times 10^{-2}$ & $4.25\times 10^{-2}$ \\
PSRJ1614-2230 & 243.65 & -22.51 & $1.21\times 10^{34}$ & 0.70 & $2.59\times 10^{-11}$ & 0.71 & $1.47\times 10^{-2}$ & $3.40\times 10^{-2}$ \\
PSRJ0312-0921 & 48.03 & -9.37 & $1.53\times 10^{34}$ & 0.82 & $5.54\times 10^{-12}$ & 0.61 & $1.36\times 10^{-2}$ & $7.27\times 10^{-3}$ \\
PSRJ0613-0200 & 93.43 & -2.01 & $1.32\times 10^{34}$ & 0.78 & $3.82\times 10^{-11}$ & 0.64 & $1.29\times 10^{-2}$ & $5.01\times 10^{-2}$ \\
PSRJ0337+1715 & 54.43 & 17.25 & $3.42\times 10^{34}$ & 1.30 &  & 0.38 & $1.20\times 10^{-2}$ &  \\
PSRJ0125-23 & 21.25 & -23.45 & $1.46\times 10^{34}$ & 0.87 &  & 0.57 & $1.14\times 10^{-2}$ &  \\
PSRJ2339-0533 & 354.91 & -5.55 & $2.32\times 10^{34}$ & 1.10 & $2.92\times 10^{-11}$ & 0.45 & $1.14\times 10^{-2}$ & $3.83\times 10^{-2}$ \\
PSRJ0621+2514 & 95.30 & 25.23 & $4.87\times 10^{34}$ & 1.64 & $4.08\times 10^{-12}$ & 0.30 & $1.07\times 10^{-2}$ & $5.35\times 10^{-3}$ \\
PSRJ1653-0158 & 253.41 & -1.98 & $1.24\times 10^{34}$ & 0.84 & $3.43\times 10^{-11}$ & 0.60 & $1.04\times 10^{-2}$ & $4.50\times 10^{-2}$ \\
PSRJ2115+5448 & 318.80 & 54.81 & $1.68\times 10^{35}$ & 3.11 & $7.02\times 10^{-12}$ & 0.16 & $1.03\times 10^{-2}$ & $9.21\times 10^{-3}$ \\
PSRJ0034-0534 & 8.59 & -5.58 & $2.97\times 10^{34}$ & 1.35 & $2.02\times 10^{-11}$ & 0.37 & $9.71\times 10^{-3}$ & $2.65\times 10^{-2}$ \\
PSRJ2042+0246 & 310.55 & 2.77 & $5.98\times 10^{33}$ & 0.64 & $5.83\times 10^{-12}$ & 0.78 & $8.67\times 10^{-3}$ & $7.65\times 10^{-3}$ \\
PSRJ1719-1438 & 259.79 & -14.63 & $1.63\times 10^{33}$ & 0.34 &  & 1.47 & $8.32\times 10^{-3}$ &  \\
PSRJ1643-1224 & 250.91 & -12.42 & $7.39\times 10^{33}$ & 0.74 &  & 0.68 & $8.01\times 10^{-3}$ &  \\
PSRJ1858-2216 & 284.57 & -22.28 & $1.13\times 10^{34}$ & 0.92 & $1.14\times 10^{-11}$ & 0.54 & $7.91\times 10^{-3}$ & $1.50\times 10^{-2}$ \\
PSRJ0251+2606 & 42.76 & 26.10 & $1.82\times 10^{34}$ & 1.17 & $4.88\times 10^{-12}$ & 0.43 & $7.90\times 10^{-3}$ & $6.40\times 10^{-3}$ \\
PSRJ0636+5128 & 6.60 & 51.48 & $5.80\times 10^{33}$ & 0.71 &  & 0.70 & $6.76\times 10^{-3}$ &  \\
PSRJ0636+5129 & 99.02 & 51.48 & $5.76\times 10^{33}$ & 0.71 &  & 0.70 & $6.71\times 10^{-3}$ &  \\
PSRJ2234+0611 & 338.60 & 6.19 & $1.04\times 10^{34}$ & 0.97 &  & 0.51 & $6.55\times 10^{-3}$ &  \\
PSRJ0248+4230 & 42.13 & 42.51 & $3.78\times 10^{34}$ & 1.85 & $1.94\times 10^{-12}$ & 0.27 & $6.54\times 10^{-3}$ & $2.55\times 10^{-3}$ \\
PSRJ1911-1114 & 287.96 & -11.24 & $1.16\times 10^{34}$ & 1.07 &  & 0.47 & $6.03\times 10^{-3}$ &  \\
PSRJ1630+3550 & 247.75 & 35.93 & $2.40\times 10^{34}$ & 1.57 &  & 0.32 & $5.80\times 10^{-3}$ &  \\
PSRJ1745-0952 & 266.29 & -9.88 & $4.97\times 10^{32}$ & 0.23 &  & 2.21 & $5.78\times 10^{-3}$ &  \\
PSRJ1012+5307 & 153.14 & 53.12 & $4.66\times 10^{33}$ & 0.70 &  & 0.71 & $5.65\times 10^{-3}$ &  \\
PSRJ0653+4706 & 103.27 & 47.11 & $7.61\times 10^{33}$ & 0.91 & $1.96\times 10^{-12}$ & 0.55 & $5.42\times 10^{-3}$ & $2.57\times 10^{-3}$ \\
PSRJ2129-0429 & 322.44 & -4.49 & $2.93\times 10^{34}$ & 1.83 & $6.81\times 10^{-12}$ & 0.27 & $5.20\times 10^{-3}$ & $8.93\times 10^{-3}$ \\
PSRJ2256-1024 & 344.23 & -10.41 & $3.71\times 10^{34}$ & 2.08 & $8.18\times 10^{-12}$ & 0.24 & $5.08\times 10^{-3}$ & $1.07\times 10^{-2}$ \\
PSRJ1630+3734 & 247.65 & 37.58 & $1.16\times 10^{34}$ & 1.19 & $5.97\times 10^{-12}$ & 0.42 & $4.89\times 10^{-3}$ & $7.83\times 10^{-3}$ \\
PSRJ2215+5135 & 333.89 & 51.59 & $6.27\times 10^{34}$ & 2.77 & $1.80\times 10^{-11}$ & 0.18 & $4.84\times 10^{-3}$ & $2.36\times 10^{-2}$ \\
PSRJ2043+1711 & 310.84 & 17.19 & $1.54\times 10^{34}$ & 1.39 & $2.85\times 10^{-11}$ & 0.36 & $4.74\times 10^{-3}$ & $3.74\times 10^{-2}$ \\
PSRJ1810+1744 & 272.66 & 17.74 & $3.84\times 10^{34}$ & 2.36 & $2.33\times 10^{-11}$ & 0.21 & $4.09\times 10^{-3}$ & $3.06\times 10^{-2}$ \\
PSRJ0406+30 & 61.64 & 30.66 & $1.84\times 10^{34}$ & 1.64 &  & 0.30 & $4.06\times 10^{-3}$ &  \\
PSRJ2017+0603 & 304.34 & 6.05 & $1.30\times 10^{34}$ & 1.40 & $3.56\times 10^{-11}$ & 0.36 & $3.94\times 10^{-3}$ & $4.67\times 10^{-2}$ \\
PSRJ2234+0944 & 338.70 & 9.74 & $1.66\times 10^{34}$ & 1.59 & $1.00\times 10^{-11}$ & 0.32 & $3.91\times 10^{-3}$ & $1.31\times 10^{-2}$ \\
PSRJ1805+0615 & 271.43 & 6.26 & $9.32\times 10^{34}$ & 3.88 & $5.27\times 10^{-12}$ & 0.13 & $3.67\times 10^{-3}$ & $6.91\times 10^{-3}$ \\
PSRJ1622-0315 & 245.75 & -3.26 & $7.93\times 10^{33}$ & 1.14 & $7.25\times 10^{-12}$ & 0.44 & $3.62\times 10^{-3}$ & $9.51\times 10^{-3}$ \\
PSRJ2055+15 & 313.95 & 15.76 & $7.92\times 10^{34}$ & 3.63 & $2.33\times 10^{-12}$ & 0.14 & $3.56\times 10^{-3}$ & $3.06\times 10^{-3}$ \\
PSRJ0751+1807 & 117.79 & 18.13 & $7.30\times 10^{33}$ & 1.11 & $1.01\times 10^{-11}$ & 0.45 & $3.52\times 10^{-3}$ & $1.33\times 10^{-2}$ \\
PSRJ1824-2452N & 18.41 & -25.87 & $1.70\times 10^{35}$ & 5.37 &  & $9.31\times 10^{-2}$ & $3.50\times 10^{-3}$ &  \\
PSRJ2302+4442 & 345.70 & 44.71 & $3.91\times 10^{33}$ & 0.86 & $3.90\times 10^{-11}$ & 0.58 & $3.12\times 10^{-3}$ & $5.12\times 10^{-2}$ \\
PSRJ0557+1550 & 89.38 & 15.84 & $1.75\times 10^{34}$ & 1.83 &  & 0.27 & $3.09\times 10^{-3}$ &  \\
\toprule\end{tabular}
\caption{List of MSPs in the stacking analysis. The hypothetical extension is $\sigma_{\rm source} =\frac{d_\mathrm{Geminga}}{d_\mathrm{Source}} \, \sigma_\mathrm{Geminga}$ for each source.}}
\end{table*}

\end{document}